\pdfoutput=1
\documentclass[11pt,twoside,a4paper,cmspaper,final,collab]{cms-tdr}
\def\svnVersion{736196b}\def\svnDate{2023/08/04}\def\cmsCernNoTag{CERN-EP-2022-199}\def\cmsCernDate{\today}\def\cmsMessage{Published in the Journal of High Energy Physics as \href{http://dx.doi.org/10.1007/JHEP07(2023)139}{\doi{10.1007/JHEP07(2023)139}.}}
\begin{document}\cmsNoteHeader{HIN-21-002}

\newlength\cmsTabSkip\setlength{\cmsTabSkip}{1ex}
\newlength\cmsNegTabSkip\setlength{\cmsNegTabSkip}{-1ex}
\newcommand{\pp}{\ensuremath{\Pp\Pp}\xspace}
\newcommand{\PbPb}{\ensuremath{\text{PbPb}}\xspace}
\newcommand{\pPb}{\ensuremath{\Pp\text{Pb}}\xspace}
\newcommand{\vsub}[2]{\ensuremath{#1_{#2}}\xspace}
\newcommand{\vsubBias}[2]{\ensuremath{#1_{#2}^{\text{bias}}}\xspace}
\newcommand{\vn}{\vsub{v}{n}}
\newcommand{\vnBias}{\vsubBias{v}{n}}
\newcommand{\vtwo}{\vsub{v}{2}}
\newcommand{\vtwoBias}{\vsubBias{v}{2}}
\newcommand{\vthree}{\vsub{v}{3}}
\newcommand{\vfour}{\vsub{v}{4}}
\newcommand{\Vn}{\vsub{V}{n\Delta}}
\newcommand{\Qtwo}{\vsub{Q}{2}}
\newcommand{\Qx}{\vsub{Q}{x}}
\newcommand{\Qy}{\vsub{Q}{y}}
\newcommand{\ptSuper}[1]{\ensuremath{\pt^{\text{#1}}}\xspace}
\newcommand{\ptCh}{\ptSuper{ch}}
\newcommand{\NText}[1]{\ensuremath{N_{\text{#1}}}\xspace}
\newcommand{\Ndijet}{\NText{dijet}}
\newcommand{\DeltaEta}{\ensuremath{\Delta\eta}\xspace}
\newcommand{\DeltaPhi}{\ensuremath{\Delta\varphi}\xspace}
\newcommand{\Sraw}{\ensuremath{S^{\text{raw}}}\xspace}
\newcommand{\xj}{\ensuremath{x_{j}}\xspace}

\cmsNoteHeader{HIN-21-002}

\title{Azimuthal anisotropy of dijet events in \PbPb collisions at \texorpdfstring{\sqrtsNN = 5.02\TeV}{sqrt(s[NN)] = 5.02 TeV}}

\date{\today}

\abstract{
  The path-length dependent parton energy loss within the dense partonic medium created in lead-lead collisions at a nucleon-nucleon center-of-mass energy of $\sqrtsNN = 5.02\TeV$ is studied by determining the azimuthal anisotropies for dijets with high transverse momentum. The data were collected by the CMS experiment in 2018 and correspond to an integrated luminosity of 1.69\nbinv. For events containing back-to-back jets, correlations in relative azimuthal angle and pseudorapidity ($\eta$) between jets and hadrons, and between two hadrons, are constructed. The anisotropies are expressed as the Fourier expansion coefficients \vn, $n = $ 2--4 of these azimuthal distributions. The dijet \vn values are extracted from long-range ($1.5 < \abs{\DeltaEta} < 2.5$) components of these correlations, which suppresses the background contributions from jet fragmentation processes. Positive dijet \vtwo values are observed which increase from central to more peripheral events, while the \vthree and \vfour values are consistent with zero within experimental uncertainties.
}

\hypersetup{
pdfauthor={CMS Collaboration},
pdftitle={Azimuthal anisotropy of jet quenching in dijet events in PbPb collisions at sqrt(s[NN]) = 5.02 TeV},
pdfsubject={CMS},
pdfkeywords={CMS, jets, flow, dijet, correlations}}

\maketitle 

\section{Introduction}

Hydrodynamic flow in relativistic heavy ion collisions is produced as initial-state geometry asymmetries transform into final-state momentum asymmetries. These asymmetries are commonly characterized by the Fourier expansion coefficients \vtwo, \vthree, \vfour, etc., of the particle azimuthal distributions. The anisotropic flow for hadrons in heavy ion collisions has been extensively studied at the BNL RHIC~\cite{PHOBOS:flow1, PHOBOS:flow2, PHOBOS:flow3, PHENIX:flow1, PHENIX:flow2, PHENIX:flow3, PHENIX:flow4, PHENIX:flow5, STAR:flow1, STAR:flow2, STAR:flow3, STAR:flow4, STAR:flow5, STAR:flow6} and at the CERN LHC~\cite{ALICE:2010pa, ALICE:2011ab, ALICE:2014wao, ALICE:2016ccg, ATLAS:2011ah, ATLAS:2012at, ATLAS:2013xzf, CMS:2012wg, CMS:2012zex, CMS:highPtV2_2012, CMS:2013wjq, CMS:2013bza, CMS:2015xmx, CMS:2017pan, CMS:2017mzx}. However, relatively few similar measurements have been done for jets~\cite{ATLAS:2013sla, ALICE:2015mda, ATLAS:2021vn}. Since partons fragmenting into high transverse momentum (\pt) jets are produced in hard processes, instead of emerging from the thermalized medium, they are not expected to ``flow" in a hydrodynamic sense. However, the jet yields can exhibit correlations with the symmetry planes in an event since the evolving parton showers experience various in-medium path lengths or medium densities as they pass through the quark-gluon plasma~\cite{jetVnTheory1, jetVnTheory2, jetVnTheory3, jetVnTheory4, jetVnTheory5}. The magnitude of these correlations depends on the details of the path-length dependent energy loss~\cite{jetVnTheory4}. In particular, the jets coplanar with the second-order symmetry plane, also known as the ``event plane", are expected to suffer less energy loss, leading to a measurable \vtwo signal. Indeed, azimuthal anisotropies of high \pt hadrons up to ${\sim}100\GeV$ have been observed \cite{CMS:highPtV2_2012, CMS:2017pan}, suggesting a path-length dependence of the parton energy loss. Performing dedicated jet azimuthal anisotropy measurements can greatly increase the accessible \pt range and give better estimates of the initial parton kinematics. The ATLAS and ALICE Collaborations have published inclusive jet \vtwo results using data for lead-lead (\PbPb) collisions at a nucleon-nucleon center-of-mass energy of $\sqrtsNN = 2.76\TeV$~\cite{ATLAS:2013sla, ALICE:2015mda}. The ATLAS collaboration subsequently extended this measurement to higher \vn harmonics at $\sqrtsNN = 5.02\TeV$~\cite{ATLAS:2021vn}. The higher \vn harmonics for jets arise from fluctuations of initial-state geometry or medium density. These studies find significant positive and centrality-dependent \vtwo values for inclusive jets. The ATLAS study~\cite{ATLAS:2021vn} also finds positive and mostly centrality-independent \vthree values for inclusive jets, while the \vfour coefficients are consistent with zero.

In this paper, we measure jet \vtwo, \vthree, and \vfour coefficients in events containing back-to-back high-\pt jets, denoted as dijet \vn coefficients, via jet-hadron correlations. Data for lead-lead collisions at $\sqrtsNN = 5.02\TeV$ with a total integrated luminosity of 1.69\nbinv~\cite{CMS:PbPbLuminosity, CMS:ppLuminosity} were collected by the CMS experiment in 2018. In contrast to previous jet \vn measurements that determine the inclusive jet \vn by the jet azimuthal correlation with the direction of maximum particle density~\cite{ATLAS:2013sla, ALICE:2015mda, ATLAS:2021vn}, here the dijet \vn is determined from jet-hadron and hadron-hadron (hereafter referred to as ``dihadron") correlations. The methodology developed for this work focuses on addressing the issue of nonflow contributions affecting the extracted harmonic coefficients. These nonflow correlations are a particular problem for high-energy jets for which, as a consequence of momentum conservation, there is a nearly back-to-back counterpart in azimuth. The ``away-side" jet fragmentation products are known to significantly contribute to the flow-like correlations. For jet-hadron correlations, these contributions from jet fragmentation are addressed taking advantage of the properties of the dijet system, and for dihadron correlations they are mitigated with a hadron \pt cut. Then, a Fourier analysis is performed on large relative pseudorapidity \DeltaEta (``long range") components of jet-hadron and dihadron correlations. To separate the jet and hadron \vn signals and extract the dijet \vn values, it is assumed that the measured jet-hadron and dihadron correlations factorize, \ie, they can be expressed as products of the jet and hadron \vn or the product of two hadron \vn values, respectively~\cite{CMS:2012wg}. This work extends the suite of experimental methods and measurements that address details of the dependence of parton energy loss on in-medium path length and medium density fluctuations. Tabulated results are provided in the HEPData record for this analysis~\cite{hepdata}.

\section{The CMS experiment}
\label{sec:cms}

The central feature of the CMS apparatus is a superconducting solenoid of 6\unit{m} internal diameter, providing a magnetic field of 3.8\unit{T}. Within the solenoid volume are a silicon pixel and strip tracker, a lead tungstate crystal electromagnetic calorimeter, and a brass and scintillator hadron calorimeter, each composed of barrel and two endcap sections. The CMS silicon tracker measures charged-particle tracks within $\abs{\eta} < 2.5$. It consists of 1856 silicon pixel and 15\,148 silicon strip detector modules. Two hadron forward (HF) steel and quartz-fiber calorimeters complement the barrel and endcap detectors, extending the calorimeter from the range $\abs{\eta} < 3.0$ provided by the barrel and endcap out to $\abs{\eta} < 5.2$. The HF calorimeters are subdivided in azimuth $(\varphi)$ into $20^\circ$ modular wedges and further segmented to form $0.175{\times}0.175$ ($\DeltaEta{\times}\DeltaPhi$) ``towers". Muons are measured in the range $\abs{\eta} < 2.4$, with detection planes located outside of the solenoid core made using three technologies: drift tubes, cathode strip chambers, and resistive plate chambers. A detailed description of the CMS detector, together with a definition of the coordinate system used and the relevant kinematic variables, can be found in Ref.~\cite{CMSexperiment}.

A particle-flow algorithm~\cite{particleFlow} using an optimized combination of information from various elements of the CMS detector is used to reconstruct leptons, photons, and charged and neutral hadrons. These are collectively referred to as particle-flow candidates. The sum of the transverse energies detected in the HF detectors $(3.0 < \abs{\eta} < 5.2)$ is used to define the event centrality~\cite{centralityDetermination} in terms of percentiles of the total inelastic hadronic cross section, with 0\% corresponding to the largest overlap of the colliding nuclei.

Events of interest are selected using a two-tiered trigger system. The first level, composed of custom hardware processors, uses information from the calorimeters and muon detectors to select events at a rate of around 100\unit{kHz}~\cite{Sirunyan:2020zal}. The second level, known as the high-level trigger, consists of a farm of processors running a version of the full event reconstruction software optimized for fast processing, and reduces the event rate to around 1\unit{kHz} before data storage~\cite{CMS:trigger2017}.

\section{Event selection}
\label{sec:eventSelection}

The events are selected using a high-level trigger that requires at least one calorimeter-based jet with $\pt > 100\GeV$. These jets are reconstructed using the anti-\kt jet clustering algorithm with a distance parameter of $R = 0.4$~\cite{antiKt}. The underlying event contribution is subtracted from the jets using an iterative method~\cite{puSubtraction} before the jet \pt is compared to the threshold. A minimum bias triggered sample is also used in the analysis as a control sample. For the minimum bias trigger, we require that on each side of the interaction point there is at least one HF tower above the readout threshold in the range of $\sim$6--12\GeV~\cite{CMS:trigger2017}.

To reduce contamination from beam-gas collisions, vertex and noise filters are applied following the example of previous analyses~\cite{CMS:2016xef}. We require that there are at least three HF towers on each side of the detector with an energy deposit of at least 3\GeV per tower. The primary vertex is required to have at least two tracks and to be reconstructed within 15\unit{cm} of the nominal interaction point in the beam direction ($z$) and within 2\unit{cm} in the transverse direction. The shapes of the clusters in the pixel detector are required to be compatible with those expected in a \PbPb collision at the vertex location. Finally, we require that there are no anomalous signal shapes or spatial distributions in the hadronic barrel and endcap calorimeter readout~\cite{CMS:hcalNoise}.

Simulated event samples are used in the analysis to correct for biases in the jet reconstruction resulting from the underlying flow modulation and to correct for jet resolution effects. These samples are produced by embedding hard jet events generated with the \PYTHIA 8.226 event generator~\cite{pythia82} with tune CP5~\cite{pythiaCP5tune} in soft underlying events (\ie, particles created from the bulk medium) simulated with the \HYDJET 1.9 event generator~\cite{hydjet}. This is denoted as the {\PYTHIA}+\HYDJET sample. The CMS detector response is simulated using the \GEANTfour toolkit~\cite{geant4}. 

Because of the large number of elementary nucleon-nucleon collisions in central \PbPb events, these events are more likely to produce jets compared to peripheral events with fewer collisions. This is taken into account in the {\PYTHIA}+\HYDJET simulation by applying a multiplicity-based weight for each event such that the charged-particle multiplicity distribution in the analysis region $\abs{\eta} < 2.4$ matches the measured distribution. An additional reweighting procedure is performed to match the position distribution of the primary vertices in the beam direction in simulation and data.

To identify high-\pt dijet events, the two jets with highest \pt in the range of $\abs{\eta} < 2$ are located. The highest \pt jet is called the ``leading jet" and is required to pass the \pt selection of $p_{\mathrm{T},1} > 120\GeV$. The second-highest \pt jet is referred to as the ``subleading jet" and is required to have $p_{\mathrm{T},2} > 50\GeV$. The azimuthal angle $\varphi$ between the leading and subleading jets is required to be $\abs{\DeltaPhi_{1,2}} > 5 \pi / 6$, ensuring that the two jets are back-to-back. Finally, both jets are required to fall within a tighter range of $\abs{\eta} < 1.3$ to ensure the most stable jet reconstruction performance and to allow for full tracker acceptance on both sides of the jets. The events containing such pairs of back-to-back jets are referred as ``dijet events".

\section{Jet and track reconstruction}
\label{sec:jetTrackReconstruction}

For this study, jets are reconstructed using the anti-\kt algorithm with a distance parameter $R = 0.4$, as implemented in the \textsc{FastJet} framework~\cite{fastJet}. Only the calorimeter information is used as an input to the anti-\kt algorithm. The underlying event contribution is subtracted from the raw jet energy using an iterative ``noise/pedestal" subtraction algorithm~\cite{puSubtraction}. First, the mean energy $\langle E_{\mathrm{T}}(\eta) \rangle$ and dispersion $\langle \sigma_{\mathrm{T}}(\eta) \rangle$ for the calorimeter cells sharing the same $\eta$ position is calculated. This determines the pedestal as a function of pseudorapidity $P(\eta) = \langle E_{\mathrm{T}}(\eta) \rangle + \langle \sigma_{\mathrm{T}}(\eta) \rangle$. Then, the pedestal values are subtracted from each calorimeter cell and jets for the next iteration step are clustered from the pedestal-subtracted calorimeter towers using the anti-\kt algorithm with $R = 0.3$. In the next iteration step, the pedestal functions are calculated again, but this time excluding all towers that are within $R = 0.5$  of any reconstructed jet with $\pt > 15\GeV$. The updated pedestal functions give the final estimate of the underlying event background. This background estimate is subtracted from the $R = 0.4$ jets and the jet energy is further calibrated using jet energy corrections calculated as a function of \pt and $\eta$ following the method described in Ref.~\cite{jetEnergyScale}.

We have chosen to reconstruct jets using only calorimeter information because this minimizes a reconstruction bias caused by the hydrodynamic flow. This bias arises from the use of the $\varphi$-averaged event energy in the underlying event subtraction. However, flow modulations lead to higher underlying event occupancies in the direction of the event plane compared to the direction perpendicular to it. This artificially enhances the \vtwo-like signal in jet-hadron correlations, since jets coplanar with the event plane have increased probability to pass the analysis selections. The same is true for higher order event planes, but with smaller effects. As calorimeters generally require higher \pt particles to produce a signal as compared to the tracker, and hydrodynamic flow is more strongly experienced by lower \pt particles, using only calorimeter information to reconstruct jets reduces the size of this bias significantly. For example, in the {\PYTHIA}+\HYDJET simulation, for the 0--10\% centrality bin, the size of this bias is $\vtwoBias \approx 0.155$ when the jets are reconstructed from particle-flow candidates, with $\vtwoBias \approx 0.079$ when only calorimeter information is used.

The track reconstruction used in \PbPb collisions is described in Ref.~\cite{trackResolution}. The charged-particle tracks used in this analysis are required to have at least 11 hits in the tracker layers and satisfy a stringent fit quality requirement, where the fit $\chi^{2}$, divided by the product of the number of fit degrees of freedom and the number of tracker layers hit, is required to be less than 0.18. To decrease the likelihood of counting nonprimary charged particles originating from secondary decay products, it is required that the distance of closest approach of a charged-particle track to at least one primary vertex in the event divided by its uncertainty is less than 3. Furthermore, it is required that the relative \pt uncertainty for the tracks is less than 10\%. Finally, in order to reduce the contribution of misreconstructed tracks with very high \pt, it is required that for tracks with $\pt > 20\GeV$, there is an associated energy deposit in the calorimeters corresponding to at least half of the track momentum. Corrections for tracking efficiency, detector acceptance, and misreconstruction rate are obtained and applied following the procedures discussed in Ref.~\cite{CMS:2016xef}.

\section{Jet-hadron and dihadron angular correlations}
\label{sec:analysisMethods}

Correlations between jets and charged particles are studied using two-dimensional distributions of the relative pseudorapidity \DeltaEta and relative azimuth \DeltaPhi of the charged particles with respect to the jet axis. These distributions are constructed by correlating each charged particle with leading and subleading jets, separately, and are normalized by the number of dijets. The analysis uses three charged particle \pt (\ptCh) bins with bin borders 0.7, 1, 2 and 3\GeV, and three centrality intervals, 0--10, 10--30, and 30--50\%. Since the majority of the measured charged particles are hadrons, the charged particles are often referred to as hadrons in this paper.

The raw correlations give the per-dijet normalized yield of leading jet-hadron or subleading jet-hadron pairs from the same event
\begin{linenomath}
\begin{equation}
  \Sraw(\DeltaEta,\DeltaPhi) = \frac{1}{\Ndijet} \frac{\rd^{2}N^{\text{same}}}{\rd\DeltaEta\rd\DeltaPhi},
  \label{eq:sameEvent}
\end{equation}
\end{linenomath}
where \Ndijet is the number of dijets satisfying the selection criteria and $N^{\text{same}}$ is the number of jet-hadron pairs. However, since the detector has limited acceptance in $\eta$, it is more likely to find jet-hadron pairs with small rather than large \DeltaEta values. Thus, the raw correlation shapes have the charged-particle yield falling rapidly towards large \DeltaEta. Detector inefficiencies can also lead to nontrivial effects on the correlation distributions. A mixed-event method, where jets and hadrons from different events are paired, is used to correct for these effects~\cite{CMS:2012wg, dijetPaper}.  By construction, such mixed-event correlations have structures due to detector and acceptance effects, but contain no physics correlations. For the mixed events, we require the primary vertex positions along the beam axis to match within 0.5\unit{cm} and the centrality percentile to be within 0.5 percentage points of those for the original data events. The charged hadrons are selected from minimum bias events to minimize jet-induced bias and to adequately capture the long-range flow correlations of the underlying event. The mixed-event pair distribution is given by
\begin{linenomath}
\begin{equation}
  ME(\DeltaEta,\DeltaPhi) = \frac{\rd^{2}N^{\text{mixed}}}{\rd\DeltaEta\rd\DeltaPhi}.
  \label{eq:mixedEvent}
\end{equation}
\end{linenomath}
The maximum of the mixed event distribution is found at $(0,0)$ since no pairs with $\DeltaEta = 0$ and $\DeltaPhi = 0$ are lost as a consequence of finite acceptance. Thus, the ratio $ME(0,0) / ME(\DeltaEta,\DeltaPhi)$ gives the normalized correction factor. Then, we construct the per-dijet associated charged-particle yield, corrected for acceptance effects, as
\begin{linenomath}
\begin{equation}
  S(\DeltaEta,\DeltaPhi) = \frac{1}{\Ndijet} \frac{\rd^{2}N}{\rd\DeltaEta\rd\DeltaPhi} = \frac{ME(0,0)}{ME(\DeltaEta,\DeltaPhi)} \Sraw(\DeltaEta,\DeltaPhi).
  \label{eq:acceptanceCorrection}
\end{equation}
\end{linenomath}

\begin{figure}[h]
  \centering{
    \includegraphics[width=0.98\linewidth]{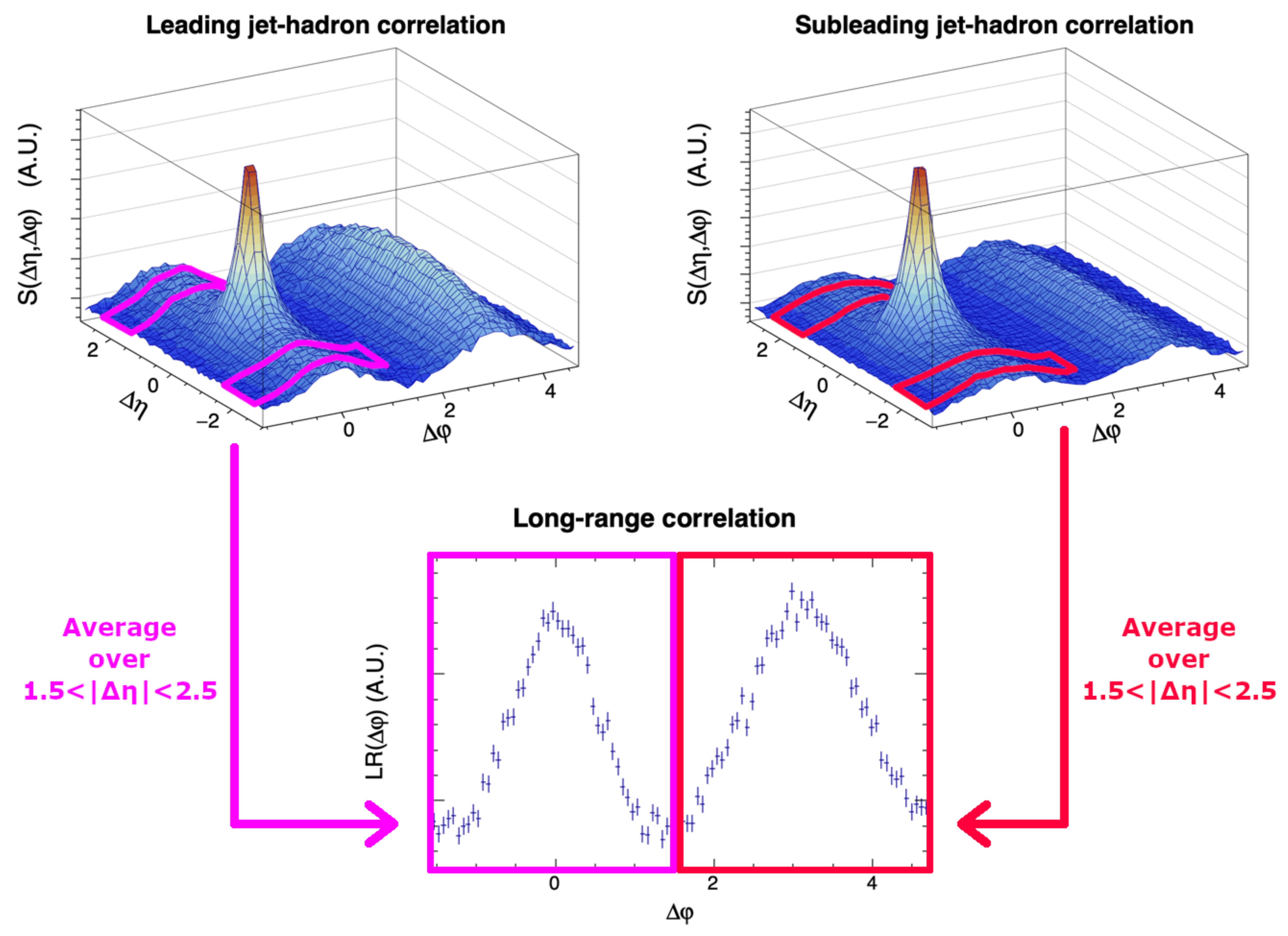} 
    \caption{Illustration on how the long-range correlation distribution is constructed. The shape of the \DeltaPhi projection corresponding to the range $1.5 < \abs{\DeltaEta} < 2.5$ is determined from both leading and subleading jet-hadron correlation distributions for $\abs{\DeltaPhi} < \pi/2$. The whole $2\pi$ range for the \DeltaPhi distribution is obtained by combining these two components.}
  \label{fig:longRangeCombinationIllustration}
  }
\end{figure}

In order to study the \vn components for the dijets, we need to separate the short-range correlations from the long-range correlations in the acceptance-corrected distribution. The short-range correlations from jet fragmentation manifest themselves as a Gaussian-like peak around $(\DeltaEta,\DeltaPhi) = (0,0)$ together with an elongated peak in \DeltaEta around $\DeltaPhi = \pi$. These are removed from the distribution by imposing selections in \DeltaEta and \DeltaPhi as illustrated in Fig.~\ref{fig:longRangeCombinationIllustration}. First, we project the \DeltaPhi distributions corresponding to the range $1.5 < \abs{\DeltaEta} < 2.5$ from both leading and subleading jet-hadron distributions. For these projections, the short-range correlation contribution to the near-side $(\abs{\DeltaPhi} < \pi/2)$ distributions is negligible, but the elongated jet peak is still present in the away-side $(\abs{\DeltaPhi} > \pi/2)$ distribution. However, in a statistical distribution, the leading and subleading jet peaks are separated by $\DeltaPhi = \pi$. It follows that for an unbiased long-range \DeltaPhi distribution $\mathrm{LR}$, we can write $\mathrm{LR}(\DeltaPhi_{\text{leading}}) = \mathrm{LR}(\pi - \DeltaPhi_{\text{subleading}})$. As the near sides of the long-range leading and subleading jet-hadron distributions have negligible bias coming from the jet peak, the long-range \DeltaPhi distribution in the entire $2\pi$ range can be found by combining the near sides of these two distributions and shifting the subleading one by $\pi$.

Once the long-range jet-hadron correlation distributions are projected on \DeltaPhi, the Fourier coefficients are found by fitting the distribution with the function
\begin{linenomath}
\begin{equation}
 f_{\mathrm{Fourier}}(\DeltaPhi) = A \left(1 + \sum_{n=1}^{4} 2 \Vn \cos(n \DeltaPhi) \right) ,
 \label{eq:fourierFit}
\end{equation}
\end{linenomath}
where $A$ is an overall normalization factor and \Vn is the fitted Fourier component of order $n$. When we fit the jet-hadron distribution, the extracted \Vn components reflect both the dijet and the hadron \vn values. 

To obtain the hadron \vn values, we also construct dihadron correlations from the same dijet events as the jet-hadron correlations. We select all charged hadrons in the analysis hadron \pt bins as trigger particles, and pair them with all other charged particles in the same \pt bin, referred to as the associated particles. We follow the same event mixing procedure to correct for the detector and acceptance effects as for the jet-hadron correlations. However, since dihadron correlations do not have jet information, the elongated jet peak cannot be removed from the away side of the correlation distribution. To mitigate the resulting background from high-\pt jet particles, we limit the hadron \pt selection with the highest $\ptCh$ studied in the analysis to $2 < \ptCh < 3\GeV$. One should notice that also for dihadron correlations, the dihadron \Vn values obtained from Eq.~\eqref{eq:fourierFit} are a mixture of trigger hadron \vn and associated hadron \vn values.

It is argued in Refs.~\cite{Luzum:vnFactorizationTheory, Alver:vnFactorizationTheory2} that in the absence of nonflow correlations, the dihadron \Vn values factorize as
\begin{linenomath}
\begin{equation}
 \Vn^{\text{dihadron}} = \vn^{\text{trigger}} \vn^{\text{associated}} ,
 \label{eq:vnFactorizationHadron}
\end{equation}
\end{linenomath}
where $\vn^{\text{trigger}}$ is the \vn of the trigger hadron and $\vn^{\text{associated}}$ is the \vn of the associated hadron. However, it is shown in Ref.~\cite{CMS:2015xmx} that \pt dependent event plane angle fluctuations can break the factorization, even without nonflow contributions. The validity of the factorization assumption is tested in Refs.~\cite{CMS:2012wg, ALICE:2011vnFactorizationTest} and shown to work with good accuracy up to $\pt^{\text{associated}} \approx 4\GeV$. Above this \pt value, the factorization assumption starts to break down, primarily because of dijet fragmentation contributions. This is reflected in the upper hadron \pt limit in this analysis, as noted earlier. The jet-hadron correlation distributions are constructed in such a way that both the near- and the away-side dijet fragmentation contributions are explicitly removed. Thus, even though jets have very high \pt particles in them, the factorization is also expected to be a good approximation for the jet-hadron correlations.

Since the same momentum range for both trigger and associated hadrons is used for dihadron correlations, the hadron \vn can be extracted by taking the square root of the fitted dihadron \Vn values. By obtaining the hadron \vn in this way, we are able to calculate dijet \vn value using the factorization assumption:
\begin{linenomath}
\begin{equation}
 \vn^{\text{dijet}} = \frac{\Vn^{\text{jet-hadron}}}{\vn^{\text{hadron}}}.
 \label{eq:vnFactorization}
\end{equation}
\end{linenomath}
Each hadron \pt bin gives us a data point for the dijet \vn in the corresponding centrality bin. Since the dijet \vn should not depend on the hadron \pt selection used to extract it, we merge all three analysis hadron \pt bins to get the dijet \vn value from one wide $0.7 < \pt < 3\GeV$ bin. The validity of the factorization assumption in Eq.~\eqref{eq:vnFactorization} was tested using {\PYTHIA}+\HYDJET simulation where we introduced a certain dijet \vtwo value at the generator level and were able to extract a consistent value following the analysis procedure.

The dijet \vn values still need to be corrected for the jet reconstruction bias. This is done by determining the dijet \vn values from {\PYTHIA}+\HYDJET simulations where jets are embedded isotropically in the azimuthally anisotropic \HYDJET background. The dijet \vn values extracted from this simulation study are, therefore, solely a result of the jet reconstruction bias. To accurately estimate the absolute amount of energy that is added to or subtracted from the jets by the azimuthal anisotropies of the underlying event, both the hadron \vtwo and the dihadron yields need to be matched simultaneously between data and simulation. This is obtained by tuning the simulation such that one of these quantities matches, and using a scaling factor for the extracted \vnBias value to take the difference found for the other into account.

For the nominal strategy, we start by matching the dihadron yields. This is achieved to a good accuracy by the multiplicity-based weighting that is applied to the {\PYTHIA}+\HYDJET simulation, as explained in Section~\ref{sec:eventSelection}. Then, we determine the hadron and dijet \vtwo values from the simulation. An additional {\PYTHIA}+\HYDJET study showed that if the dihadron yield is kept constant and the hadron \vtwo is varied by a certain percentage, the dijet \vtwo value also changes by the same percentage. This means that the scaling factor for \vnBias that accounts for the difference in the simulated and observed hadron \vtwo can be directly obtained from the data-to-simulation hadron \vtwo ratio. The jet reconstruction bias is then corrected by subtracting the scaled \vnBias value obtained this way from the raw data dijet \vn value.

The second strategy used to evaluate systematic uncertainties matches the hadron \vtwo, and then applies a scaling factor to take into account the differences in dihadron yields. As a starting point for this strategy, we use a {\PYTHIA}+\HYDJET simulation, where instead of using the nominal multiplicity-based weighting to take into account the larger number of nucleon-nucleon collisions in central events, we use a centrality-based weighting scheme. In this scheme, we weight the centrality distribution determined from the HF calorimeters to match between data and simulation. Then, we check the underlying event energy density in random cones to see which centrality range in the simulation corresponds to similar energy densities in the data. The best match is found when the nominal centrality definition in the simulation is shifted 4 percentage points upwards. In this case, for example, the 0--10\% centrality bin in data is matched with the 4--14\% centrality bin in the {\PYTHIA}+\HYDJET simulation. After the centrality distributions are matched, we apply an event shape engineering method presented in Ref.~\cite{eventShapeEngineering} to match the hadron \vtwo between simulation and data. It is shown in Ref.~\cite{eventShapeEngineering} that the elliptic flow and the magnitude of the second-order flow vector \Qtwo normalized by the square root of event multiplicity are correlated. Thus, a selection based on this variable in the {\PYTHIA}+\HYDJET simulation event-by-event can be done to control the extracted hadron \vtwo value. The \Qtwo-vector magnitude is defined as
\begin{linenomath}
\begin{equation}
 \Qtwo = \sqrt{\smash[b]{\Qx^{2} + \Qy^{2}}},
\end{equation}
\end{linenomath}
where
\begin{linenomath}
\begin{equation}
 \Qx = \sum_{i} \cos(2 \varphi_{i}),\; \Qy = \sum_{i} \sin(2 \varphi_{i}) .
\label{eq:QxQy}
\end{equation}
\end{linenomath}
Only particles from the \HYDJET part of the simulation in $\abs{\eta} < 0.75$ and $\pt < 3\GeV$ are used. Using a \Qtwo-vector selection where hadron \vtwo values in data and simulation match, the dijet \vn values are determined from the {\PYTHIA}+\HYDJET simulation. We have found in the previously described additional {\PYTHIA}+\HYDJET study that when hadron \vtwo is kept the same, changing multiplicity does not affect the dijet \vtwo linearly. Instead, we found a multiplicity-dependent function which allows us to calculate different scaling factors for each centrality bin. Thus, the dijet \vnBias values obtained after applying the \Qtwo-vector selection need to be scaled by the ratio of dihadron yields in data and simulation times this centrality dependent scaling factor. As before, the obtained dijet \vnBias values are subtracted from the raw data dijet \vn values to get the final corrected dijet \vn results.

\section{Systematic uncertainties}
\label{sec:systematics}

The following sources of systematic uncertainty are considered in this analysis:
\begin{itemize}
  \item \emph{Acceptance correction}. Since jet correlations are small-angle correlations, and long-range correlations only depend on \DeltaPhi, the \DeltaEta distribution at $\abs{\eta} > 1.5$ should be uniform. To evaluate possible deviations from the uniformity that might arise from an imperfect acceptance correction, the analysis is repeated, extracting the long-range correlation distribution only from the negative ($-2.5<\DeltaEta<-1.5$) or positive ($1.5<\DeltaEta<2.5$) sides of \DeltaEta. The larger difference from the nominal result is assigned as a systematic uncertainty.
  \item \emph{Long-range extraction}. Uncertainties resulting from the long-range correlation distribution are determined by projecting the \DeltaPhi distributions from two parts of the extraction region, $1.5<\abs{\DeltaEta}<2.0$ and $2.0<\abs{\DeltaEta}<2.5$. The larger difference from the nominal result is assigned as a systematic uncertainty.
  \item \emph{Jet angular resolution}. The uncertainty in the jet angular resolution is estimated by determining the resolution in the {\PYTHIA}+\HYDJET simulation by comparing the $\eta$ and $\varphi$ values for jet axes before and after detector effects are considered. The nominal $\eta$ and $\varphi$ values for jet axes in data are then varied by Gaussian distributions with widths based on the resolutions found in the simulation. The difference in results with and without the additional Gaussian dispersion is taken as the uncertainty.
  \item \emph{Dijet bias in dihadron correlations}. It is possible that the dijet selection changes the hadron \vn with respect to minimum bias events. To check for this effect, we repeat the dihadron correlation measurement using a minimum bias data sample, and use the difference from the nominal results as an uncertainty.
    \item \emph{Jet energy scale}. The related uncertainties are estimated by varying the jet energy corrections within their uncertainties and seeing how these changes affect the final correlations. The jet energy correction procedure is detailed in Ref.~\cite{jetEnergyScale}.
  \item \emph{Jet energy resolution}. This uncertainty is estimated by comparing the nominal results with the ones obtained by adding a Gaussian spread to the nominal jet energies, as a function of jet \pt, such that the jet energy resolution estimated from the simulation is worsened by 20\%. The value of 20\% is determined by comparing dijet momentum balance $\xj = \pt^{\text{subleading}} / \pt^{\text{leading}}$ distributions in peripheral 50--70\% and 70--90\% bins between data and {\PYTHIA}+\HYDJET simulation. The jet energy resolution in the simulation is worsened by different amounts, and comparing the shapes of the resulting \xj distributions to data, it is seen that the maximal difference between jet energy resolutions in data and simulation is 20\%.
  \item \emph{Tracking efficiency}. The tracking-related uncertainties are estimated by repeating the analysis without any tracking corrections.
  \item \emph{Jet reconstruction bias correction}. There are several sources of uncertainty related to the jet reconstruction bias correction. First, there is uncertainty on the dijet \vn values determined from the {\PYTHIA}+\HYDJET simulation. The dijet \vn values from the simulation in each centrality bin are extracted by performing a constant fit to the results from different hadron \pt bins up to $\ptCh = 4\GeV$. The uncertainty of this fit is included in the uncertainty of the jet reconstruction bias correction.
  
  Second, we compare the dijet \vn results obtained using two different matching strategies between simulation and data to determine the correction. Both of these are described in detail in the end of Section~\ref{sec:analysisMethods}.
  
  Third, the quark/gluon jet fraction in the {\PYTHIA}+\HYDJET simulation can be different from data, affecting the jet reconstruction bias correction. The potential difference is estimated to be less than 25\% using a template fit to the multiplicity distribution of particle-flow candidates within the jet cone in the data. Then, the uncertainty is estimated by varying the quark/gluon jet fraction in simulation by this amount.

\end{itemize}

The total systematic uncertainties are obtained by adding all the individual components together in quadrature. The relative contributions from different sources are listed in Table~\ref{tab:systematicUncertaintiesVn} for the different dijet \vn harmonics. It can be seen from this table that the dominant source of uncertainty in most of the analysis bins arises from jet reconstruction. The jet reconstruction bias uncertainty is generally larger for higher \vn harmonics. While the simulated sample size is the same, the higher \vn coefficients have smaller signal sizes. Thus, they cannot be determined as accurately as the \vtwo values for the correction.

\begin{table}[htbp]
\centering{
  \topcaption{The breakdown of different sources of systematic uncertainty for dijet \vn, separately for the three centrality bins considered in the analysis.}
  \label{tab:systematicUncertaintiesVn}
  \begin{tabular}{ccccc}
  \vn  & Source & 0--10\% &  10--30\%  & 30--50\%  \\ 
    \hline
   \multirow{9}{*}{\vtwo} & Acceptance correction & $0.002$ & ${<}0.001$ & $0.001$ \\
    & Long-range extraction & $0.003$ & $0.003$ & $0.002$ \\
    & Jet angle resolution & ${<}0.001$ & ${<}0.001$ & $0.001$ \\
    & Jet reconstruction bias & $0.008$ & $0.003$ & $0.006$ \\
    & Dijet bias for dihadron & $0.002$ & $0.001$ & $0.001$ \\
    & Tracking & ${<}0.001$ & $0.001$ & ${<}0.001$ \\
    & Jet energy scale & $0.002$ & $0.001$ & $0.002$ \\
    & Jet energy resolution & $0.004$ & $0.003$ & $0.002$ \\[\cmsTabSkip]
    & Total for \vtwo & $0.010$ & $0.005$ & $0.007$ \\[\cmsTabSkip]
    \hline \\[\cmsNegTabSkip]
    \multirow{9}{*}{\vthree} & Acceptance correction & ${<}0.001$ & $0.001$ & $0.002$ \\
    & Long-range extraction & $0.002$ & $0.001$ & $0.006$ \\
    & Jet angle resolution & $0.001$ & $0.001$ & $0.001$ \\
    & Jet reconstruction bias & $0.005$ & $0.016$ & $0.016$ \\
    & Dijet bias for dihadron & ${<}0.001$ & $0.001$ & $0.001$ \\
    & Tracking & ${<}0.001$ & ${<}0.001$ & $0.001$ \\
    & Jet energy scale & $0.001$ & $0.001$ & $0.004$ \\
    & Jet energy resolution & $0.003$ & $0.001$ & $0.001$ \\[\cmsTabSkip]
    & Total for \vthree & $0.006$ & $0.017$ & $0.017$ \\[\cmsTabSkip] 
    \hline \\[\cmsNegTabSkip] 
    \multirow{9}{*}{\vfour} & Acceptance correction & $0.003$ & $0.002$ & $0.005$ \\
    & Long-range extraction & $0.003$ & $0.003$ & $0.001$ \\
    & Jet angle resolution & $0.001$ & ${<}0.001$ & ${<}0.001$ \\
    & Jet reconstruction bias & $0.018$ & $0.016$ & $0.026$ \\
    & Dijet bias for dihadron & ${<}0.001$ & ${<}0.001$ & ${<}0.001$ \\
    & Tracking & ${<}0.001$ & ${<}0.001$ & $0.002$ \\
    & Jet energy scale & $0.003$ & $0.001$ & $0.003$ \\
    & Jet energy resolution & $0.002$ & $0.003$ & $0.002$ \\[\cmsTabSkip]
    & Total for \vfour & $0.019$ & $0.017$ & $0.026$ \\    
  \end{tabular}
}
\end{table}

\section{Results}
\label{sec:results}

\begin{figure}[ht]
  \centering{
    \includegraphics[width=0.98\linewidth]{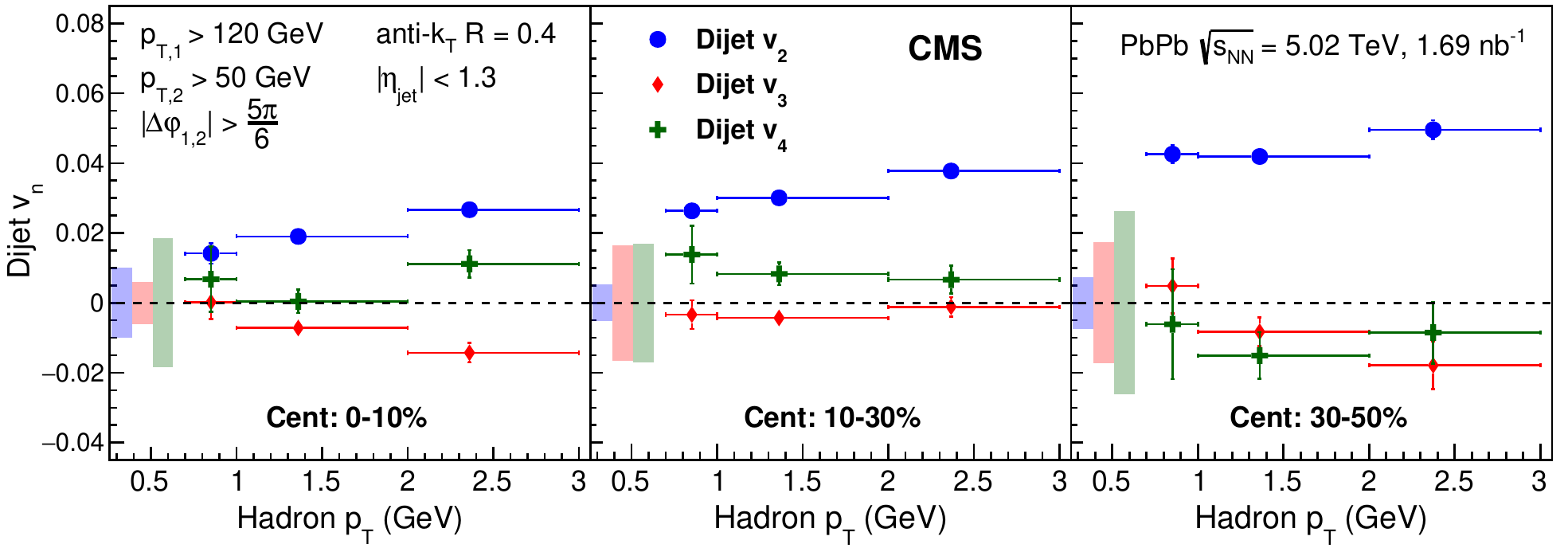} 
    \caption{The dijet \vn data points factorized using different associated hadron \pt bins for 0--10\% (left), 10--30\% (middle), and 30--50\% (right) centrality bins. The data points are corrected for the jet reconstruction bias effects. The vertical bars represent statistical uncertainties, while the \pt-independent systematic uncertainties are plotted as shaded areas on the left side of the panels.}
  \label{fig:jetVnFitIllustration}
  }
\end{figure}

The extracted dijet \vtwo, \vthree, and \vfour values in different hadron \pt bins are presented in Fig.~\ref{fig:jetVnFitIllustration}. All results shown in this figure are corrected for the jet reconstruction bias effects. Some dependence of the dijet \vn values on the reference particle \pt is observed, which is consistent with the expectation of possible factorization breaking by residual back-to-back correlations.

\begin{figure}[ht]
  \centering{
    \includegraphics[width=0.98\linewidth]{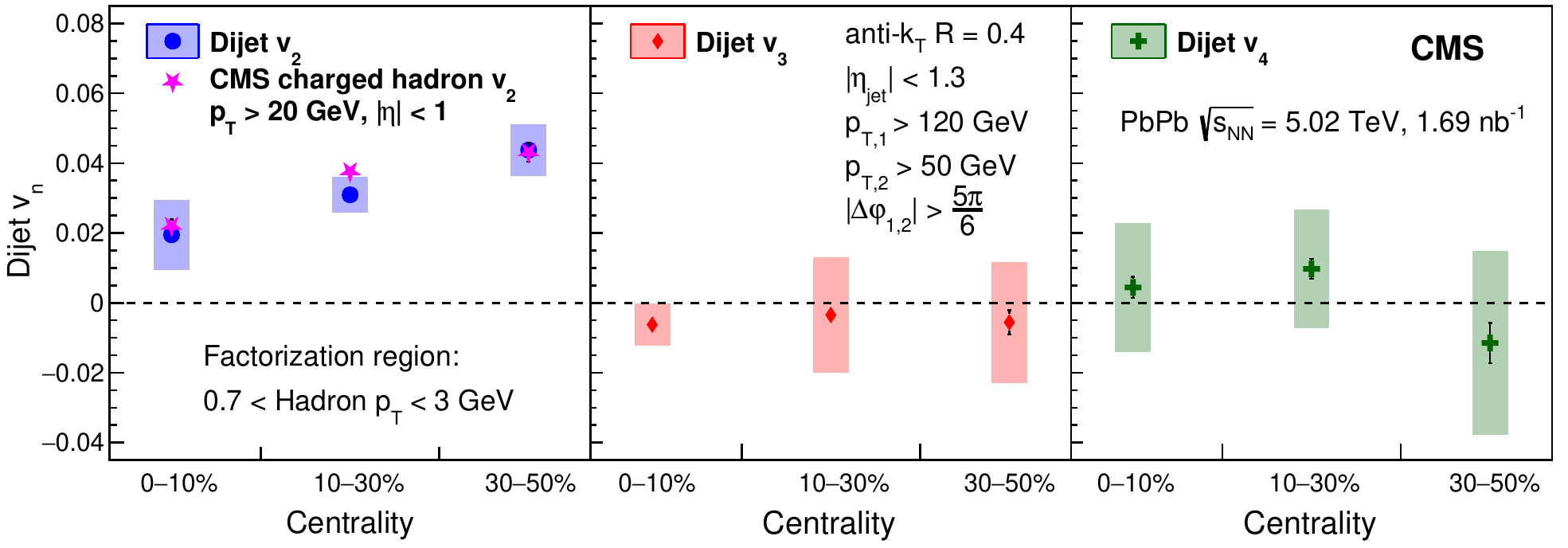} 
    \caption{Dijet \vtwo (left), \vthree (middle), and \vfour (right) results presented as functions of centrality. The dijet \vtwo results are compared to CMS high-\pt hadron \vtwo results from Ref.~\cite{CMS:2017pan}. The shaded areas represent systematic uncertainties and the vertical bars are the statistical uncertainties.}
  \label{fig:finalResults}
  }
\end{figure}

The centrality dependence of the dijet \vn averaged over the reference particle \pt range of 0.7--3\GeV is presented in Fig.~\ref{fig:finalResults}. The dijet \vtwo measurements show positive values, indicating more jets observed coplanar with the event plane compared to the perpendicular direction. Since jets coplanar with event plane traverse less medium, these jets suffer less energy loss on average compared to those in the perpendicular direction. Thus, they are more likely to pass the analysis cuts, leading to the observed \vtwo signal. The dijet \vtwo magnitude is found to increase toward more peripheral collisions up to 30--50\%, which is expected based on the increasing eccentricity of the collision overlap region. The current measurements are compared with previous CMS results on high-\pt hadron \vtwo from Ref.~\cite{CMS:2017pan}. Since that earlier work used finer centrality bins, the high-\pt hadron \vtwo values plotted in Fig.~\ref{fig:finalResults} are compiled by first combining the centrality bins to match the ones used in this analysis, weighting each centrality bin by the number of events. Then, all hadron \vtwo points above $20\GeV$ are fitted with a constant to define a value corresponding to contributions from the jet fragmentation. We observe that the measured dijet \vtwo values are consistent with the high-\pt hadron \vtwo values within the uncertainties, and similar values are also observed by the ATLAS collaboration in Ref.~\cite{ATLAS:2021vn}.

The dijet \vthree and \vfour results in the middle and right panels of Fig.~\ref{fig:finalResults} are found to be compatible with zero within experimental uncertainties in each centrality bin. Recent theory calculations for high-\pt hadrons~\cite{jetVnTheory4} and $R = 0.2$ inclusive jets~\cite{jetVnTheory5} predict jet \vthree values that are positive and less than 0.005 in magnitude. At this level, the precision of the current measurement is not sufficient to probe the impact of the fluctuations in the initial-state geometry and medium density on the dijet azimuthal distributions. The results for the dijet \vthree values are consistent with the CMS high-\pt hadron measurements in Ref.~\cite{CMS:2017pan}, which are also compatible with zero. In contrast, the recent ATLAS results show positive inclusive jet \vthree for $R=0.2$ jets above 71\GeV~\cite{ATLAS:2021vn}. However, the ATLAS analysis is made with different selection criteria (lower jet \pt and smaller distance parameter) and the ATLAS inclusive jet and CMS dijet populations are different, so the two results should not be directly compared.

\section{Summary}
\label{sec:conclusions}

The Fourier coefficients \vtwo, \vthree, and \vfour are determined for jets from events containing back-to-back jets (``dijet \vn'') in lead-lead collisions at $\sqrtsNN = 5.02\TeV$. The jet-hadron correlation technique used for this measurement has been developed to unambiguously separate jet fragmentation-related contributions from the long-range correlations due to the in-medium path length and medium density dependencies of parton energy loss.

The dijet \vtwo values are found to be positive, meaning that more jets are observed coplanar with the event plane than perpendicular to this plane. The dijet \vtwo values increase with increasing eccentricity of the initial collision region, from about 2.0\% in the 0--10\% centrality bin to about 4.4\% in the 30--50\% centrality bin. These results are qualitatively consistent with expectations from a path-length dependence of in-medium energy loss. For all measured centrality bins, the dijet \vthree and \vfour values are consistent with zero within experimental uncertainties. However,
the current results do not have the precision needed to probe the effects of the initial-state geometry and medium density fluctuations on the dijet azimuthal distributions. The measured dijet \vn values provide valuable input to a more precise and quantitative description of the partonic energy loss in the quark-gluon plasma.

\begin{acknowledgments}
We congratulate our colleagues in the CERN accelerator departments for the excellent performance of the LHC and thank the technical and administrative staffs at CERN and at other CMS institutes for their contributions to the success of the CMS effort. In addition, we gratefully acknowledge the computing centers and personnel of the Worldwide LHC Computing Grid and other centers for delivering so effectively the computing infrastructure essential to our analyses. Finally, we acknowledge the enduring support for the construction and operation of the LHC, the CMS detector, and the supporting computing infrastructure provided by the following funding agencies: BMBWF and FWF (Austria); FNRS and FWO (Belgium); CNPq, CAPES, FAPERJ, FAPERGS, and FAPESP (Brazil); MES and BNSF (Bulgaria); CERN; CAS, MoST, and NSFC (China); MINCIENCIAS (Colombia); MSES and CSF (Croatia); RIF (Cyprus); SENESCYT (Ecuador); MoER, ERC PUT and ERDF (Estonia); Academy of Finland, MEC, and HIP (Finland); CEA and CNRS/IN2P3 (France); BMBF, DFG, and HGF (Germany); GSRI (Greece); NKFIH (Hungary); DAE and DST (India); IPM (Iran); SFI (Ireland); INFN (Italy); MSIP and NRF (Republic of Korea); MES (Latvia); LAS (Lithuania); MOE and UM (Malaysia); BUAP, CINVESTAV, CONACYT, LNS, SEP, and UASLP-FAI (Mexico); MOS (Montenegro); MBIE (New Zealand); PAEC (Pakistan); MES and NSC (Poland); FCT (Portugal); MESTD (Serbia); MCIN/AEI and PCTI (Spain); MOSTR (Sri Lanka); Swiss Funding Agencies (Switzerland); MST (Taipei); MHESI and NSTDA (Thailand); TUBITAK and TENMAK (Turkey); NASU (Ukraine); STFC (United Kingdom); DOE and NSF (USA).
  
\hyphenation{Rachada-pisek} Individuals have received support from the Marie-Curie program and the European Research Council and Horizon 2020 Grant, contract Nos.\ 675440, 724704, 752730, 758316, 765710, 824093, 884104, and COST Action CA16108 (European Union); the Leventis Foundation; the Alfred P.\ Sloan Foundation; the Alexander von Humboldt Foundation; the Belgian Federal Science Policy Office; the Fonds pour la Formation \`a la Recherche dans l'Industrie et dans l'Agriculture (FRIA-Belgium); the Agentschap voor Innovatie door Wetenschap en Technologie (IWT-Belgium); the F.R.S.-FNRS and FWO (Belgium) under the ``Excellence of Science -- EOS" -- be.h project n.\ 30820817; the Beijing Municipal Science \& Technology Commission, No. Z191100007219010; the Ministry of Education, Youth and Sports (MEYS) of the Czech Republic; the Hellenic Foundation for Research and Innovation (HFRI), Project Number 2288 (Greece); the Deutsche Forschungsgemeinschaft (DFG), under Germany's Excellence Strategy -- EXC 2121 ``Quantum Universe" -- 390833306, and under project number 400140256 - GRK2497; the Hungarian Academy of Sciences, the New National Excellence Program - \'UNKP, the NKFIH research grants K 124845, K 124850, K 128713, K 128786, K 129058, K 131991, K 133046, K 138136, K 143460, K 143477, 2020-2.2.1-ED-2021-00181, and TKP2021-NKTA-64 (Hungary); the Council of Science and Industrial Research, India; the Latvian Council of Science; the Ministry of Education and Science, project no. 2022/WK/14, and the National Science Center, contracts Opus 2021/41/B/ST2/01369 and 2021/43/B/ST2/01552 (Poland); the Funda\c{c}\~ao para a Ci\^encia e a Tecnologia, grant CEECIND/01334/2018 (Portugal); the National Priorities Research Program by Qatar National Research Fund; MCIN/AEI/10.13039/501100011033, ERDF ``a way of making Europe", and the Programa Estatal de Fomento de la Investigaci{\'o}n Cient{\'i}fica y T{\'e}cnica de Excelencia Mar\'{\i}a de Maeztu, grant MDM-2017-0765 and Programa Severo Ochoa del Principado de Asturias (Spain); the Chulalongkorn Academic into Its 2nd Century Project Advancement Project, and the National Science, Research and Innovation Fund via the Program Management Unit for Human Resources \& Institutional Development, Research and Innovation, grant B05F650021 (Thailand); the Kavli Foundation; the Nvidia Corporation; the SuperMicro Corporation; the Welch Foundation, contract C-1845; and the Weston Havens Foundation (USA).
\end{acknowledgments}
\bibliography{auto_generated}

\cleardoublepage \appendix\section{The CMS Collaboration \label{app:collab}}\begin{sloppypar}\hyphenpenalty=5000\widowpenalty=500\clubpenalty=5000
\cmsinstitute{Yerevan Physics Institute, Yerevan, Armenia}
{\tolerance=6000
A.~Tumasyan\cmsAuthorMark{1}\cmsorcid{0009-0000-0684-6742}
\par}
\cmsinstitute{Institut f\"{u}r Hochenergiephysik, Vienna, Austria}
{\tolerance=6000
W.~Adam\cmsorcid{0000-0001-9099-4341}, J.W.~Andrejkovic, T.~Bergauer\cmsorcid{0000-0002-5786-0293}, S.~Chatterjee\cmsorcid{0000-0003-2660-0349}, K.~Damanakis\cmsorcid{0000-0001-5389-2872}, M.~Dragicevic\cmsorcid{0000-0003-1967-6783}, A.~Escalante~Del~Valle\cmsorcid{0000-0002-9702-6359}, P.S.~Hussain\cmsorcid{0000-0002-4825-5278}, M.~Jeitler\cmsAuthorMark{2}\cmsorcid{0000-0002-5141-9560}, N.~Krammer\cmsorcid{0000-0002-0548-0985}, L.~Lechner\cmsorcid{0000-0002-3065-1141}, D.~Liko\cmsorcid{0000-0002-3380-473X}, I.~Mikulec\cmsorcid{0000-0003-0385-2746}, P.~Paulitsch, F.M.~Pitters, J.~Schieck\cmsAuthorMark{2}\cmsorcid{0000-0002-1058-8093}, R.~Sch\"{o}fbeck\cmsorcid{0000-0002-2332-8784}, D.~Schwarz\cmsorcid{0000-0002-3821-7331}, M.~Sonawane\cmsorcid{0000-0003-0510-7010}, S.~Templ\cmsorcid{0000-0003-3137-5692}, W.~Waltenberger\cmsorcid{0000-0002-6215-7228}, C.-E.~Wulz\cmsAuthorMark{2}\cmsorcid{0000-0001-9226-5812}
% ...
\par}
\vskip\cmsinstskip
\dag:~Deceased\\
$^{1}$Also at Yerevan State University, Yerevan, Armenia\\
$^{2}$Also at TU Wien, Vienna, Austria\\
% ...
\end{sloppypar}
\end{document}